**Title**: Open Source Software for Efficient and Transparent Reviews

**Authors**: Rens van de Schoot*[1]; Jonathan de Bruin[2]; Raoul Schram[2]; Parisa Zahedi[2], Jan de Boer[3], Felix Weijdema[3], Bianca Kramer[3], Martijn Huijts[4], Maarten Hoogerwerf[2], Gerbrich Ferdinands[1], Albert Harkema[1], Joukje Willemsen[1], Yongchao Ma[1], Qixiang Fang[1], Sybren Hindriks[1], Lars Tummers[5], Daniel L. Oberski[1,6]

[1] Department of Methodology and Statistics, Faculty of Social and Behavioral Sciences, Utrecht University, The Netherlands

[2] Department of Research and Data Management Services, Information Technology Services, Utrecht University, The Netherlands

[3] Utrecht University Library, Utrecht University, The Netherlands

[4] Department of Test and Quality Services, Information Technology Services, Utrecht University, The Netherlands

[5] School of Governance, Faculty of Law, Economics and Governance, Utrecht University, The Netherlands

[6] Department of Biostatistics, Data management and Data Science, Julius Center, University Medical Center Utrecht, The Netherlands

**Corresponding author**: Rens van de Schoot: Department of Methods and Statistics, Utrecht University, P.O. Box 80.140, 3508TC, Utrecht, The Netherlands; Tel.: +31 302534468; E-mail address: a.g.j.vandeschoot@uu.nl.

**Funding**: This project was funded by the Innovation Fund for IT in Research Projects, Utrecht University, The Netherlands.

**Abstract**

To help researchers conduct a systematic review or meta-analysis as efficiently and transparently as possible, we designed a tool (`ASReview`) to accelerate the step of screening titles and abstracts. For many tasks - including but not limited to systematic reviews and meta-analyses - the scientific literature needs to be checked systematically. Currently, scholars and practitioners screen thousands of studies by hand to determine which studies to include in their review or meta-analysis. This is error prone and inefficient because of extremely imbalanced data: only a fraction of the screened studies is relevant. The future of systematic reviewing will be an interaction with machine learning algorithms to deal with the enormous increase of available text. We therefore developed an open source machine learning-aided pipeline applying active learning: `ASReview`. We demonstrate by means of simulation studies that `ASReview` can yield far more efficient reviewing than manual reviewing, while providing high quality. Furthermore, we describe the options of the free and open source research software and present the results from user experience tests. We invite the community to contribute to open source projects such as our own that provide measurable and reproducible improvements over current practice.

**Keywords**

Active Learning, Machine Learning, Researcher-in-the-loop, Open Science, Systematic reviewing

## Main Text

With the emergence of online publishing, the number of scientific papers on many topics is skyrocketing.[1] All these textual data present opportunities to scholars and practitioners, while simultaneously confronting them with new challenges. To develop comprehensive overviews of the relevant topics, scholars often develop systematic reviews and meta-analyses.[2] The process entails several explicit and, ideally, reproducible steps, including identifying all likely relevant publications in a standardized way, extracting data from eligible studies, and synthesizing the results. Systematic reviews differ from traditional literature reviews in that they are more replicable and transparent.[3,4] Such systematic overviews of literature on a specific topic are pivotal not only for scholars, but also for clinicians, policy makers, journalists, and, ultimately, the general public.[5–7]

Given that screening the entire research literature on a given topic is too labor intensive, scholars often develop quite narrow searches. Developing a search strategy for a systematic review is an iterative process aimed at balancing recall and precision;[8,9] that is, including as many potentially relevant studies as possible, while at the same time limiting the total number of studies retrieved. Often, the vast number of publications in the field of study may lead to a relatively precise search, with the risk of missing relevant studies. The process of systematic reviewing is error-prone and extremely time intensive.[10] In fact, if a field's literature is growing faster than the amount of time available for systematic reviews, adequate manual review of this field becomes impossible.[11]

To aid researchers the rapidly evolving field of machine learning (ML) has allowed the development of software tools that assist in developing systematic reviews.[11–14] It offers approaches to overcome the manual and time-consuming screening of large numbers of studies by prioritizing relevant studies using active learning.[15] Active learning is a type of machine learning in which a model can choose the data points (e.g., records obtained from a systematic search) it would like to learn from, and thereby drastically reducing the total number of records that require manual screening.[16–18] In most, so-called, Human-in-the-Loop (HITL)[19] machine learning applications, the interaction between the machine learning algorithm and the human is used to train a model with a minimum number of labeling tasks. Unique for systematic reviewing is that not only all relevant records (i.e., titles and abstracts) should be seen by a researcher, but also an extremely diverse range of concepts needs to be learned, thereby requiring flexibility in the modeling approach as well as careful error evaluation.[11] In the case of systematic reviewing, the algorithm(s) are interactively optimized for finding the most relevant records, instead of finding the most accurate model. Therefore, the term Researcher-In-The-Loop (RITL) was introduced[20] as a special case of HITL with three unique components: (1) The primary output of the process is a selection of the records, not a trained machine learning model, (2) All records in the relevant selection are seen by a human at the end of the process[21], (3) The use-case requires a reproducible workflow and complete transparency is required.[22].

Existing tools implementing such an active learning cycle for systematic reviewing are described in Table 1, see the Appendix for an overview of all the software we considered (note this list was based on a review of software tools[12] ). However, existing tools have two main drawbacks. First, many are closed source applications with black box algorithms. This is problematic as transparency and data ownership are  essential in the era of open science[22]. Second, to our knowledge, existing tools lack the necessary flexibility to deal with the large range of possible concepts to be learned by a screening machine. For example, in systematic reviews, the optimal type of classifier will depend on variable parameters, such as the proportion of relevant publications in the initial search and the complexity of the inclusion criteria used by the researcher.[23] For this reason any successful system must allow for a wide range of classifier types. Benchmark testing is crucial to understand the real-world performance of any ML-aided system, but currently, such benchmark options are mostly lacking.

<< TABLE 1 >>

In this paper, we present an open source ML-aided pipeline with active learning for systematic reviews called `ASReview`. The goal of `ASReview` is to help scholars and practitioners to get an overview of the most relevant records for their work as efficiently as possible, while being transparent in the process. The open, free and ready-to-use software `ASReview` addresses all concerns mentioned above: it is open source, uses active learning, allows multiple ML-models. It also has a benchmark mode which is especially useful for comparing and designing algorithms. Furthermore, it is intended to be easily extensible, allowing third parties to add modules that enhance the pipeline. Although we focus this paper on systematic reviews, ASReview can handle any text source.

In what follows, we first present the pipeline for manual versus ML-aided systematic reviews. Subsequently, we present how `ASReview` has been set up, and how `ASReview` can be used in different workflows by presenting several real-world use cases. Then, we present the results of simulations that benchmark performance, and present the results of a series of user-experience tests. Last, we discuss future directions.

**Pipeline for manual and ML-aided systematic reviews**

Traditionally, the pipeline of a systematic review without active learning starts with researchers doing a comprehensive search in multiple databases[24], using free text words as well as controlled vocabulary to retrieve potentially relevant references. The researcher then typically verifies that key papers they expect to find are indeed included in the search results. The researcher downloads a file with records containing the text to be screened. In the case of systematic reviewing it contains the titles and abstracts, and potentially other metadata like authors, journal, DOI, of potentially relevant references into a reference manager. Ideally, two or more researchers then screen the records' titles and abstracts based on eligibility criteria established beforehand.[4] After all records have been screened, the full texts of the potentially relevant records are read to analyze which will be ultimately included in the review. Most records are excluded in the title and abstract phase. Typically, only a small fraction of the records belong to the relevant class, making title and abstract screening an important bottleneck in systematic reviewing process [25]. For instance, a recent study analyzed 10,115 records, and excluded 9,847 after title and abstract screening, a drop of more than 95%.[26] Therefore, `ASReview` focuses on this labor-intensive step.

The research pipeline of `ASReview` is depicted in Figure 1. The researcher starts with a search exactly as described above, and subsequently uploads a file containing the records (i.e. metadata containing the text of the titles and abstracts) into the software. Prior knowledge is then selected which is used for training of the first model and presenting the first record to the researcher. Because screening is a binary classification problem, the reviewer must select at least one key record to include *and* exclude based on background knowledge. More prior knowledge may result in improved efficiency of the active learning process.

<< FIGURE 1>>

Based on the prior knowledge, a machine learning classifier is trained to predict study relevance (labels) from a representation of the record containing text (feature space). In order to prevent "authority bias" in the inclusions, we have purposefully chosen *not* to include an author name or citation network representation in the feature space. In the active learning cycle, the software presents one new record to be screened and labeled (1 - "relevant" vs. 0 – "irrelevant") by the user. The user's binary label is subsequently used to train a new model, after which a new record is presented to the user. This cycle continues up to a certain user-specified stopping criterion has been reached. The user now has a file with (1) records labeled as either relevant or irrelevant and (2) unlabeled records ordered from most to least probable to be relevant as predicted by the current model. This setup helps to move through a large database much quicker than in the manual process, while, at the same time, the decision process remains transparent.

### Software implementation: `ASReview`

The source code[27] of `ASReview` is available open source under an Apache-2.0 license, including documentation[28]. Compiled and packaged versions of the software are available on the Python Package Index[29] or Docker Hub[30]. The free and ready-to-use software `ASReview` implements an 'oracle', a 'simulation' and an 'exploration' mode. The oracle mode is used to perform a systematic review with interaction by the user. The simulation mode is used for simulation of the ASReview performance on existing systematic reviews. The exploration mode can be used for teaching purposes and includes several pre-loaded labeled datasets.

The oracle mode presents records to the researcher, and the researcher classifies these. Multiple file formats are supported: (1) RIS files are used by digital libraries, like IEEE Xplore, Scopus and ScienceDirect. Citation managers Mendeley, RefWorks, Zotero, and EndNote support the RIS format as well. (2) Tabular datasets with extensions .csv, .xlsx, and .xls. CSV files should be comma separated and UTF-8 encoded. For CSV files, the software accepts a set of predetermined labels in line with the ones used in RIS files. Each record in the dataset should hold metadata on, for example, a paper. Mandatory metadata is text and can for example be titles or abstracts from scientific papers. If available, both are used to train the model, but at least one is needed. An advanced option is available which splits the title and abstracts in the feature extraction step and weights the two feature matrices independently (for TF-IDF only). Other metadata such as author, date, URL, DOI, and keywords are optional but not used for training the models. When using ASReview in simulation or exploration mode, an additional binary variable to indicate historical labeling decisions is required. This column, which is automatically detected, can also be used in the oracle mode as background knowledge for prior selection of relevant papers before entering the active learning cycle. If not available the user has to select at least one relevant record which can be identified by searching the pool of records. Also, at least one irrelevant record should be identified; the software allows to search for specific records or presents random records which are most likely to be irrelevant due to the extremely imbalanced data.

The software has a simple yet extensible default model: a Naive Bayes classifier, TF-IDF feature extraction, Dynamic Resampling balance strategy[31], and certainty-based sampling[17,32] for the query strategy. These defaults were chosen based on their consistently high performance in benchmark experiments across several datasets[31]. Moreover, the low computation time of these default settings makes them attractive in applications, given that the software should be able to run locally. Users can change the settings, shown in Table 2, and technical details are described in our documentation[28]. Users can also add their own classifiers, feature extraction techniques, query strategies and balance strategies.

<< TABLE 2 >>

`ASReview` has a number of implemented features (see Table 2). First, there are several classifiers available: (1) naive Bayes, (2) support vector machines, (3) logistic regression, (4) neural networks, (5) random forests, (6) LSTM-base which consists of an embedding layer, an LSTM layer with one output, a dense layer, and a single sigmoid output node, and (7) LSTM-pool which consists of an embedding layer, an LSTM layer with many outputs, a max pooling layer, and single sigmoid output node. Feature extraction techniques available are Doc2Vec,[33] embedding with IDF or TF-IDF[34] (the default is unigram, with the option to run n-grams, while other parameters are set to the defaults of Scikit-learn[35]), and sBERT.[36] The available query strategies for the active learning part are (1) Random selection, ignoring model assigned probabilities, (2) Uncertainty-based sampling which chooses the most uncertain record according to the model (i.e. closest to 0.5 probability), (3) Certainty-based sampling ("Max" in `ASReview`) which chooses the record most likely to be included according to the model, and (4) Mixed sampling which uses a combination of random and certainty-based sampling.

There are several balance strategies that rebalance and reorder the training data. This is necessary, because the data is typically extremely imbalanced and therefore we have implemented the following balance strategies: (1) Full sampling which uses all the labeled records, (2) Undersampling the irrelevant records, so that the included and excluded records are in some particular ratio (closer to one), and (3) "Dynamic Resampling", a novel method similar to undersampling in that it decreases the imbalance of the training data[31]. However, in Dynamic Resampling, the number of irrelevant records is decreased, whereas the number of relevant records is increased by duplication such that the total number of records in the training data remains the same. The ratio between relevant and irrelevant records is not fixed over interactions, but dynamically updated, depending on the number of labeled records, the total number of records and the ratio between relevant and irrelevant records. Details on all the described algorithms can be found in the code and documentation referred to above.

By default, ASReview converts the records' texts into a document-term matrix, terms are converted to lowercase, and no stop words are removed as default (but this can be changed). Because the document-term matrix is identical in each iteration of the active learning cycle, it is generated in advance of model training, and stored in the (active learning) state file. The indexed records can easily be requested from the document-term matrix in the state file. Internally, records are identified by their row number in the input dataset. In “oracle mode”, the record that is selected to be classified is retrieved from the state file and the record text and other metadata (such as title and abstract) are retrieved from the original dataset (from file or computer memory). `ASReview` can run on your local computer, or a (self-hosted) local or remote server. Data - all records and their labels - remain on the users’ computer. Data ownership and confidentiality is crucial, and no data is processed or used in any way by third parties. This stands in distinction with some of the existing systems, as shown in the last column of Table 1.

**Real world use-cases and high-level function descriptions**

Below we highlight a number of real-world use cases and high-level function descriptions for using the pipeline of `ASReview`.

`ASReview` can be integrated in classic systematic reviews or meta-analyses. Such reviews or meta-analyses entail several explicit and reproducible steps, as outlined in the PRISMA guidelines.[4] Scholars identify all likely relevant publications in a standardized way, screen retrieved publications to select eligible studies based on defined eligibility criteria, extract data from eligible studies and synthesize the results. `ASReview` fits in this process, particularly in the abstract screening phase. `ASReview` does not replace the initial step of collecting all potentially relevant studies. As such, results from `ASReview` depend on the quality of the initial search process, including selection of databases[24] and construction of comprehensive searches using keywords and controlled vocabulary. However, `ASReview` can be used to broaden the scope of the search, by keyword expansion or by omitting limitation in the search query, resulting in a higher number of initial papers to limit the risk of missing relevant papers during the search part (i.e., more focus on recall instead of precision).

Also, when analyzing very large literature streams, many reviewers nowadays move towards meta-reviews, that is, systematic reviews of systematic reviews.[37] This can be problematic as the various reviews included could use different eligibility criteria and therefore are not always directly comparable. Because of the efficiency of `ASReview`, scholars using the tool could conduct the study by analyzing the papers directly instead of using the systematic reviews. Furthermore, ASReview supports the rapid update of a systematic review. The included papers from the initial review are used to train the machine learning model before screening of the updated set of papers starts. This allows the researcher to quickly screen the updated set of papers based on decisions made in the initial run.

As an example case, let us look at the current literature on COVID-19 and the coronavirus. An enormous number of papers are being published on COVID-19 and the coronavirus. It is very time consuming to manually find relevant papers, for example to develop treatment guidelines. This is especially problematic as urgent overviews are required. Medical guidelines rely on comprehensive systematic reviews, but the medical literature is growing at breakneck pace, and the quality of the research is not universally adequate for summarization into policy.[38] Such reviews must entail adequate protocols with explicit and reproducible steps, including identifying all potentially relevant papers, extracting data from eligible studies, assessing potential for bias, and synthesizing the results into medical guidelines. Researchers need to screen (tens of) thousands of COVID-19 related studies by hand to find relevant papers to include in their overview. Using ASReview, this can be done far more efficiently by selecting key papers that match their (COVID-19) research question in the first step; this should start the active learning cycle and lead to the most relevant COVID-19 papers for their research question being presented next. Therefore, a plug-in was developed for `ASReview`[39] containing three databases which are updated automatically whenever a new version is released by the owners of the data: (1) The Cord19 database, developed by the Allen Institute for AI, with over all publications on COVID-19 and other coronavirus research (e.g. SARS, MERS, etc.) from PubMed Central, the WHO COVID-19 database of publications, the preprint servers bioRxiv and medRxiv and papers contributed by specific publishers[40]. The CORD-19 dataset is updated daily by the Allen Institute for AI and updated also daily in the plugin. (2) In addition to the full dataset, we construct automatically a daily subset of the database with studies published after December 1st, 2019 to search for relevant papers published during the COVID-19 crisis. (3) A separate dataset of COVID-19 related preprints, containing metadata of preprints from over 15 preprints servers across disciplines, published since January 1, 2020.[41] The preprint dataset is updated weekly by the maintainers and then automatically updated in ASReview as well. As this dataset is not readily available to researchers through regular search engines (e.g. PubMed), its inclusion in ASReview provided added value to researchers interested in COVID-19 research, especially if they want a quick way to screen preprints specifically.

**Simulation study**

To evaluate the performance of `ASReview` on a labeled dataset, users can employ the simulation mode. As an example, we ran simulations based on four labeled datasets with version 0.7.2 of `ASReview`. All scripts to reproduce the results in this paper can be found on Zenodo (doi:10.5281/zenodo.4024122)[42] and the results are available at OSF (doi:10.17605/OSF.IO/2JKD6)[43].

<< FIGURE 2 >>

***Datasets.*** First, we analyzed the performance for a study systematically describing studies that performed viral Metagenomic Next-Generation Sequencing (mNGS) in common livestock such as cattle, small ruminants, poultry, and pigs.[44] Studies were retrieved from Embase (n = 1,806), Medline (n = 1,384), Cochrane Central (n = 1), Web of Science (n = 977), and Google Scholar (n = 200, the top relevant references). After deduplication this led to 2,481 studies obtained in the initial search, of which 120 inclusions (4.84%).

A second simulation study was performed on the results for a systematic review of studies on fault prediction in software engineering[45]. Studies were obtained from ACM Digital Library, IEEExplore and the ISI Web of Science. Additionally, a snowballing strategy and a manual search were conducted, accumulating to 8,911 publications of which 104 were included in the systematic review (1.2%).

A third simulation study was performed on a review of longitudinal studies that applied unsupervised machine learning techniques on longitudinal data of self-reported symptoms of posttraumatic stress assessed after trauma exposure[46,47] 5,782 studies were obtained by searching Pubmed, Embase, PsychInfo, and Scopus, and through a snowballing strategy in which both the references and the citation of the included papers were screened. Thirty-eight studies were included in the review (0.66%).

A fourth simulation study was performed on the results for a systematic review on the efficacy of Angiotensin-converting enzyme (ACE) inhibitors, from a study collecting various systematic review datasets from the medical sciences[15]. The collection is a subset of 2,544 publications from the TREC 2004 Genomics Track document corpus[48]. This is a static subset from all MEDLINE records from 1994 through 2003, which allows for replicability of results. Forty-one publications were included in the review (1.6%).

***Performance Metrics.*** We evaluated the four datasets using three performance metrics. First, we assess the "Work Saved over Sampling" (WSS). WSS is the percentage reduction in the number of records needed to screen that is achieved by using the program instead of screening records at random. WSS is measured at a given level of recall of relevant records, for example 95%, indicating the work reduction in screening effort at the cost of failing to detect 5% of the relevant records. For some researchers it is essential that all relevant literature on the topic is retrieved; this entails that the recall should be 100% (i.e., WSS@100%). Note that to be sure to detect 100% of relevant records, all records need to be screened, therefore leading to no time savings.  We also propose the amount of Relevant References Found after having screened the first 10% of the records, RRF10%. This is a useful metric for getting a quick overview of the relevant literature.

***Results.*** For every dataset, 15 runs were performed with one random inclusion and one random exclusion, see Figure 2. The classical review performance with randomly found inclusions is shown by the dashed line. The average work saved over sampling at 95% recall for `ASReview` is 83% and ranges from 67% to 92%. Hence, 95% of the eligible studies will be found after screening between only 8% to 33% of the studies. Furthermore, the number of relevant abstracts found after reading 10% of the abstracts ranges from 70% to 100%. In short, our software would have saved many hours of work.

**Usability Testing (UX-Testing)**

We conducted a series of user experience tests to learn from end users how they experience the software and implement it in their workflow. The study was approved by the Ethics Committee of the Faculty of Social and Behavioral Sciences of Utrecht University (ID 20-104).

***Unstructured Interviews.*** The first user experience (UX) test, carried out in December 2019, was conducted with an academic research team in a substantive research field (public administration and organizational science) that has conducted various systematic reviews and meta-analyses. It was composed of three university professors (ranging from assistant to full) and three PhD candidates. In one 3.5-hour session, the participants used the software and provided feedback via unstructured interviews and group discussions. The goal was to provide feedback on installing the software and testing the performance on their own data. After these sessions we prioritized the feedback in a meeting with the ASReview team which resulted in the release of v0.4[49] and v0.6[50]. An overview of all releases can be found on GitHub[27].

A second UX-test was conducted with four experienced researchers developing medical guidelines based on classical systematic reviews, and two experienced reviewers working at a pharmaceutical non-profit organization who work on updating reviews with new data. In four sessions, held in February-March 2020, these users tested the software following our testing protocol . After each session we implemented the feedback provided by the experts and asked them to review the software again. The main feedback was about how to upload datasets and select prior papers. Their feedback resulted in the release of v0.7[51] and v0.9[52].

***Systematic UX-Test.*** In May 2020, we conducted a systematic UX-test. Two groups of users were distinguished: an unexperienced group and an experienced user who already used ASReview. Due to the COVID-19 lockdown the usability tests were conducted via video calling where one person gave instructions to the participant and one person observed, called human-moderated remote testing[53]. During the tests, one person (SH) asked the questions and helped the participant with the tasks, the other person observed and made notes, a user experience professional at the IT-department of Utrecht University (MH).

To analyze the notes, thematic analysis was used, which is a method to analyze data by dividing the information in subjects that all have a different meaning[54] using the software Nvivo 12[55]. When something went wrong the text was coded as “showstopper”. When something did not go smoothly the text was coded as “doubtful”. When something went well the subject was coded as “superb”. The features the participants requested for future versions of the ASReview tool were discussed with the lead engineer of the ASReview team and were submitted to GitHub as issues or feature requests.

The answers to the quantitative questions can be found at the Open Science Framework[56]. The participants (N=11) rated the tool with a grade of 7.9 (SD = 0.9) on a scale from one to ten (Table 2). The unexperienced users on average rated the tool with an 8.0 (SD= 1.1, N=6). The experienced user on average rated the tool with a 7.8 (SD= 0.9, N=5). The participants described the usability test with words such as “helpful”, “accessible”, “fun”, “clear” and “obvious”.

The UX-tests resulted in the new release v0.10[57], v0.10.1[58] and the major release v0.11[59], which is a major revision of the GUI. The documentation has been upgraded to make installing and launching `ASReview` more straightforward. We made setting up the project, selecting a dataset and finding prior knowledge is more intuitive and flexible. In addition, we added a project dashboard with information on your progress and advanced settings.

***Continuous Input via Open Source Community.*** Finally, the `ASReview` development team receives continuous feedback from the open science community about, among other things, the user experience. In every new release we implement features listed by our users. Recurring UX-tests are done to keep up with the needs of users and improve the value of the tool.

## Conclusion

To help researchers conduct a systematic review or meta-analysis as efficiently and transparently as possible, we designed a system to accelerate the step of screening titles and abstracts. Our system uses active learning to train an ML model that predicts relevance from texts using a limited number of labeled examples. The classifier, feature extraction technique, balance strategy, and active learning query strategy are flexible. We provide an open source software implementation, `ASReview` with state-of-the-art systems across a wide range of real-world systematic reviewing applications. Based on our experiments, `ASReview` provides defaults on its parameters which exhibited good performance on average across the applications we examined. However, we stress that in practical applications, these defaults should be carefully examined; for this purpose, the software provides a simulation mode to users. We encourage users and developers to perform further evaluation of the proposed approach in their application, and to take advantage of the project's open source nature by contributing additional developments.

Drawbacks of ML-based screening systems, including our own, remain. First, while the active learning step greatly reduces the number of papers that must be screened, it also prevents a straightforward evaluation of the system's error rates without further onerous labeling. Providing users with an accurate estimate of the system's error rate in the application at hand is therefore a pressing open problem. Second, while, as argued above, the use of such systems is not limited in principle to reviewing, to our knowledge no empirical benchmarks of actual performance in these other situations yet exist. Third, ML-based screening systems automate the *screening* step only; while the screening step is time-consuming and a good target for automation, it is just one part of a much larger process, including the initial search, data extraction, coding for risk of bias, summarizing results, etc. While some other work, similar to our own, has looked at (semi-)automating some of these steps in isolation[60,61], to our knowledge the field is still far removed from an integrated system that would truly automate the review process while guaranteeing the quality of the produced evidence synthesis. Integrating the various tools that are currently under development to aid the systematic reviewing pipeline is therefore a worthwhile topic for future development.

Possible future research could also focus on the performance of identifying full text articles with different document length and domain-specific terminologies or even other types of text, such as newspaper articles and court cases. When the selection of prior knowledge is not possible based on expert knowledge, alternative methods could be explored. For example, unsupervised learning or pseudo-labeling algorithms could be used to improve training[62,63]. In addition, as the NLP community pushes forward the state of the art in feature extraction methods, these are easily added to our system as well. In all cases, performance benefits should be carefully evaluated using benchmarks for the task at hand. To this end, common benchmark challenges should be constructed that allow for an even comparison of the various tools now available. To facilitate such a benchmark, we have constructed a repository of publicly available systematic reviewing datasets[64].

The future of systematic reviewing will be an interaction with machine learning algorithms to deal with the enormous increase of available text. We invite the community to contribute to open source projects such as our own, as well as to common benchmark challenges, so that we can provide measurable and reproducible improvement over current practice.

**Acknowledgement**: We would like to thank the Utrecht University Library, focus area *Applied Data Science*, and departments of Information and Technology Services, Test and Quality Services, and Methodology & Statistics, for their support. We also want to thank all researchers who shared data, participated in our user experience tests or who gave us feedback on `ASReview` in other ways. Furthermore, we would like to thank the editors and reviewers for providing constructive feedback.

**Author Contributions**:

RvdS and DO originally designed the project, with later input from LT. JdBr is the lead engineer, software architect and supervises the code base on GitHub. RS coded the algorithms and simulation studies. PZ coded the very first version and created Figure 2 together with BK. JdBo, FW, MH and BK developed the systematic review pipeline. MH is leading the UX-tests and was supported by SH. MH developed the architecture of the produced (meta)data. GF conducted the simulation study together with RS. AH performed the literature search comparing the different tools together with GF. JW designed all the artwork and helped with formatting the manuscript. YM and QF are responsible for the pre-processing of the metadata under the supervision of JdBr. RvdS, DO and LT have written the paper with input from all authors. Each co-author has written parts of the manuscript.

**Competing Interests statement**: There is no competing interest.

**Data and Code Availability statement:** All code to reproduce the results described in this paper can be found on Zenodo (doi:10.5281/zenodo.4024122)[42] and the results are available at the Open Science Framework (doi:10.17605/OSF.IO/2JKD6)[43]. All code for the software `ASReview` is available under an Apache-2.0 license *(doi:10.5281/zenodo.3345592)*[27], is maintained on GitHub[65], and includes documentation (doi:10.5281/zenodo.4287120)[28]. The answers to the quantitative questions of the UX-test can be found at the Open Science Framework (OSF.IO/7PQNM)[56].

**Table 1.** In this table we provide an overview of those tools that implemented active learning and describe what machine learning algorithms have been implemented, which active learning features are available and information about privacy policy. As a starting point we used the systematic review[12] describing ML-aided software tools for systematic reviewing. In Table A1 in the Appendix we provide an overview of all tools found by Harrison et al. and indicate which tools implemented machine learning and/or active learning and are open source. Note that we added FASTREAD, RobotAnalyst and ASReview to the overview which were not described by Harrison et al.

| Name | Machine learning algorithms | Active learning features | Privacy Policy |
|---|---|---|---|
| Abstrackr [66] | **Classifier**: SVM.<br><br>**Model inputs**: User-provided keywords (relevant/irrelevant with degree of confidence); citations.<br><br>**Feature extraction:** TF-IDF**.**<br><br>**Label options**: Relevant; borderline; irrelevant. | **Query strategy**: Uncertainty-based; Certainty-based; Random sampling.<br><br>**Balance strategy:** Aggressive undersampling.<br><br>**Active learning starts after:** A reasonable representation of the minority class has been labeled (Wallace et al., 2010)**.**<br><br>**Retraining:** Asynchronous.<br><br>**Stopping:** When the model predicts none of the remaining abstracts to be relevant. | **<u>GDPR Notification</u>**:<br>*"We do not have a limit on how long we retain your account information and/or data."*<br>*"We do not share any information with third parties."* |
| ASReview [27] | **Classifier**: NB; SVM; DNN; LR; LSTM-base; LSTM-pool; RF.<br><br>**Model inputs:** piece of text, for example title and abstract.<br><br>**Feature extraction:** Doc2Vec; embedding-IDF, TF-IDF, sBERT.<br><br>**Label options:** Relevant; irrelevant. | **Query strategy**: Uncertainty-based; Certainty-based; Random sampling; Mixed Sampling<br><br>**Balance strategy:** Simple (no balancing); Dynamic Resampling (double and triple); Undersampling.<br><br>**Active learning starts after:** One label.<br><br>**Retraining:** Asynchronous<br><br>**Stopping:** Is currently left to the reviewer. | Software does not have access to user data, because the program runs locally. |
| Colandr[67] | **Classifier**: SVM with SGD learning.<br><br>**Model inputs:** User-provided key terms and citation (abstract, title, keywords).<br><br>**Feature extraction:** Word2Vec.<br><br>**Label options:I**nclude; exclude. | **Query strategy**: Certainty-based<br><br>**Balance strategy:** Reweighting.<br><br>**Active learning starts after:** 100 inclusions and 100 exclusions.<br><br>**Retraining:** Every 30 abstracts.<br><br>**Stopping:** Is left to the reviewer. | No terms and conditions available.<br><br>The Colandr team was contacted and they ensured the user can remove data any time. In the future, user data will be used to improve Colandr but only if granted permission from the project owner |
| FASTREAD[68] | **Classifier**: SVM.<br><br>**Model inputs:** Title and abstract.<br><br>**Feature extraction:** TF-IDF.<br><br>**Label options:** Relevant; irrelevant | **Query strategy**: Uncertainty sampling; Certainty sampling. Users are allowed to switch between active learning types after 30 inclusions.<br><br>**Balance strategy:** Mix of weighting and aggressive undersampling.<br><br>**Active learning starts after:** One relevant abstract is retrieved (through querying random abstracts).<br><br>**Retraining:** every 10 abstracts.<br><br>**Stopping:** The number of relevant abstracts is estimated by semi-supervised learning. | Software does not have access to user data, because the program runs locally. |
| Rayyan[69] | **Classifier**: SVM.<br><br>**Model inputs:** User-provided key terms and citation (title and abstract).<br><br>**Feature extraction:** Unigrams, bigrams**,** MeSH terms<br><br>**Label options:** Include; exclude; maybe. | **Query strategy**: Rayyan predicts a relevancy of a citation on a 5-star scale. The user can order citations by their predicted relevancy.<br><br>**Balance strategy:** Unknown.<br><br>**Active learning starts after:** Unknown.<br><br>**Retraining:** Unknown; *'as the user is labelling citations*.<br><br>**Stopping:** When there are no more citations to be labeled or when the model can no longer be improved. | **Rayyan Terms of Service:**<br>3.1: *"Rayyan, may use any User data and information to evaluate and improve its performance and expand its services."*<br>3.4: *"This Agreement is governed by the laws of the State of Qatar. By accessing this Rayyan website you consent to these terms and conditions and to the exclusive jurisdiction of the Qatar courts in all disputes arising out of such access."*<br>9.2.2: *"Rayyan does not own User Content. The User retains the copyright of their Content. …"* |
| RobotAnalyst[70] | **Classifier**: SVM.<br><br>**Model inputs:** Title; abstract; topic model proportions.<br><br>**Feature extraction:** TF-IDF L2 normalised (title); bow for abstract; LDA for topic model proportions.<br><br>**Label options:** Included; excluded; undecided. | **Query strategy**: Uncertainty-based; certainty-based.<br><br>**Balance strategy:** None.<br><br>**Active learning starts after:** A manually labeled “initial batch” of abstracts, randomly sampled or obtained through a focused search.<br><br>**Retraining:** When to retrain is left to the user. Possible after every labeled citation.<br><br>**Stopping:** Is left to the reviewer, however at least a sequence of excluded citations is necessary. | Not available. |

Notes: Machine learning = kind of Machine learning model used; Active Learning = how Active Learning is implemented; Privacy Policy = if available, quotes from privacy policy are given to indicate possible concerns. SVM - Support Vector Machine; TF-IDF - Term Frequency - Inverse Document Frequency; NB - Naïve Bayes; DNN - Dense Neural Network; LR - Logistic Regression; LSTM - Long short-term memory; RF - Random Forests; Doc2Vec - Document to Vector; Embedding-IDF - Embedding Inverse Document Frequency; S-BERT - Sentence Bidirectional Encoder Representations from Transformers; SGD - Stochastic Gradient Descent; Word2Vec - Words to Vector; Bow - bag of words; LDA - Latent Dirichlet Allocation; GDPR - General Data Protection Regulation; MeSH - Medical Subject Headings.

**Table 2.** Implemented classifiers, feature extraction techniques, query strategies and balance strategies available in ASReview. Note that, not all combinations are possible. For example, the naive Bayes classifier cannot handle a feature matrix with negative values, so that this classifier cannot be combined with Doc2Vec; LSTM-base and LSTM-pool classifiers exclusively work with embeddingLSTM feature extraction and vice versa. Technical details are described in our documentation[28].

| Classifier | Feature extraction | Query strategy | Balance strategy |
|---|---|---|---|
| Naive Bayes (*default*)<br>Support vector machine<br>Neural network<br>Logistic regression<br>LSTM-base<br>LSTM-pool<br>Random forests | TF-IDF (*default*)<br>embedding-IDF<br>Sentence BERT<br>Doc2Vec<br>EmbeddingLSTM | Certainty-based sampling (*default*)<br>Uncertainty-based sampling<br>Random sampling<br>Mixed sampling (e.g., 5% random sampling / 95% certainty-based) | "Dynamic resampling" (double and triple) (double=*default*)<br>Undersampling<br>Simple (no balancing) |

### Figures

**Figure 1.** ML-aided pipeline for Automated Systematic Review (`ASReview`). The symbols indicate whether the action is taken by a human, a computer, or if both options are available.

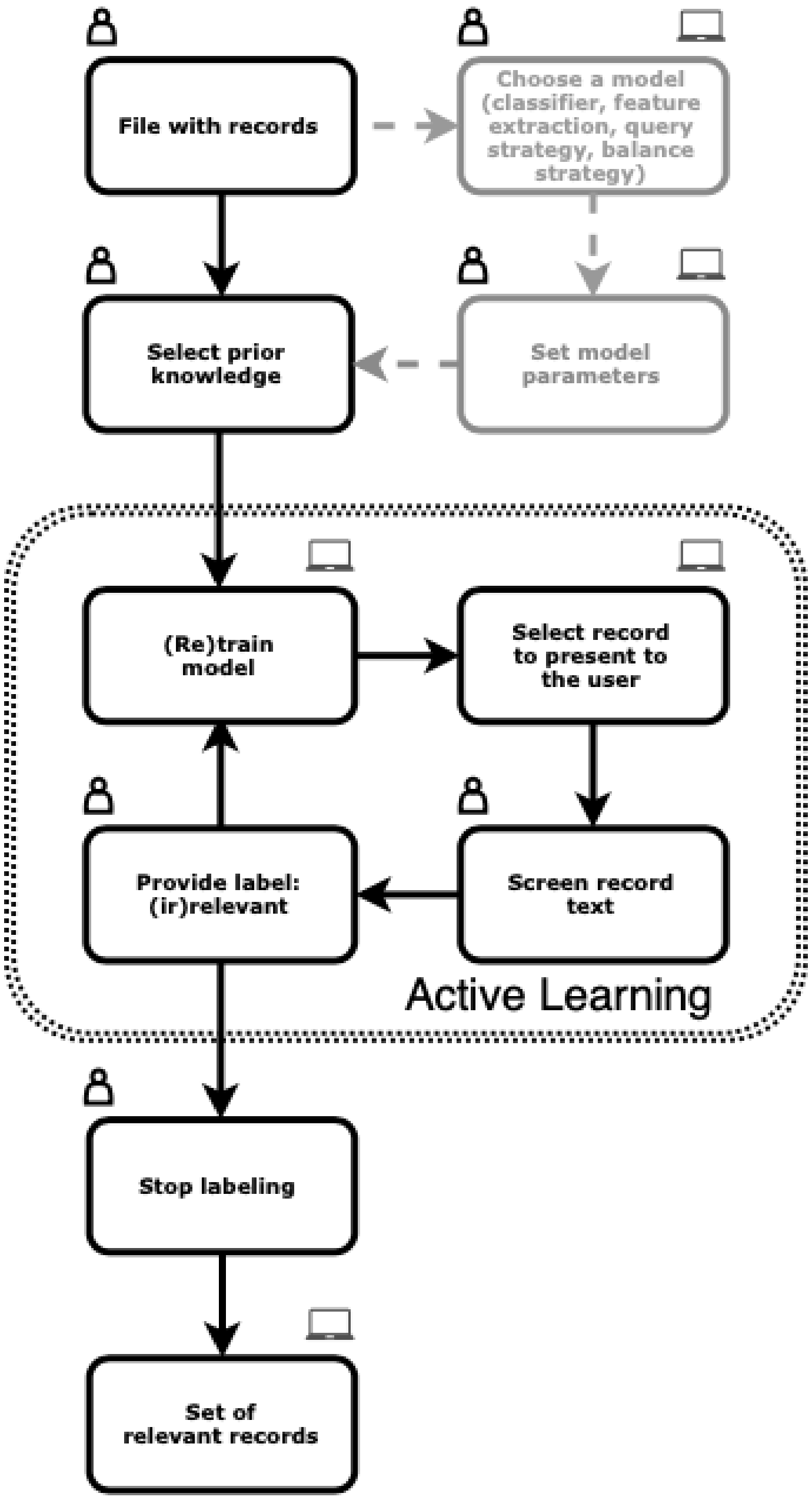

**Figure 2.** Results of the simulation study for the results for a study systematically review studies that performed viral Metagenomic Next-Generation Sequencing (mNGS) in common livestock (**a**), results for a systematic review of studies on fault prediction in software engineering (**b**), results for longitudinal studies that applied unsupervised machine learning techniques on longitudinal data of self-reported symptoms of posttraumatic stress assessed after trauma exposure (**c**), and results for a systematic review on the efficacy of Angiotensin-converting enzyme (ACE) inhibitors (**d**). For every dataset, 15 runs, shown with separate lines, were performed with only one random inclusion and one random exclusion. The classical review performance with randomly found inclusions is shown by the dashed line.

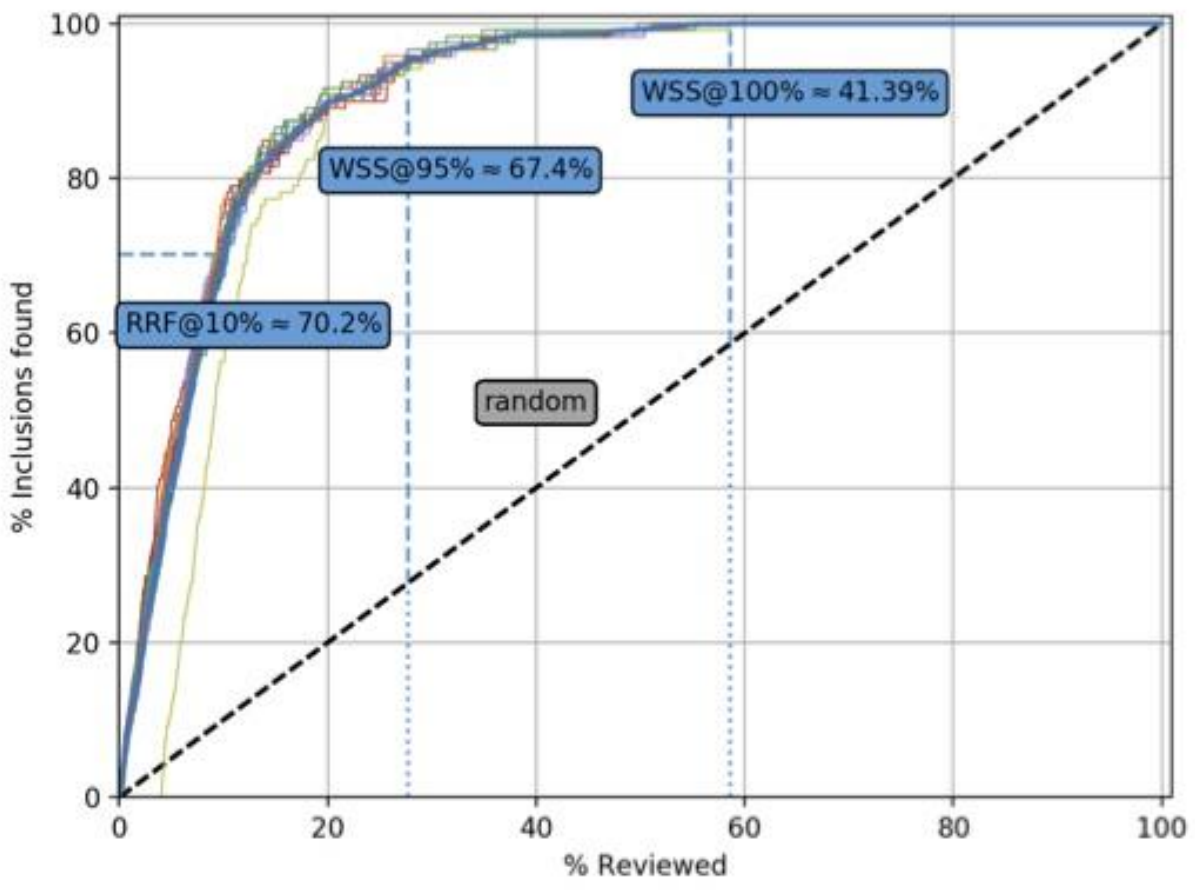

(a)

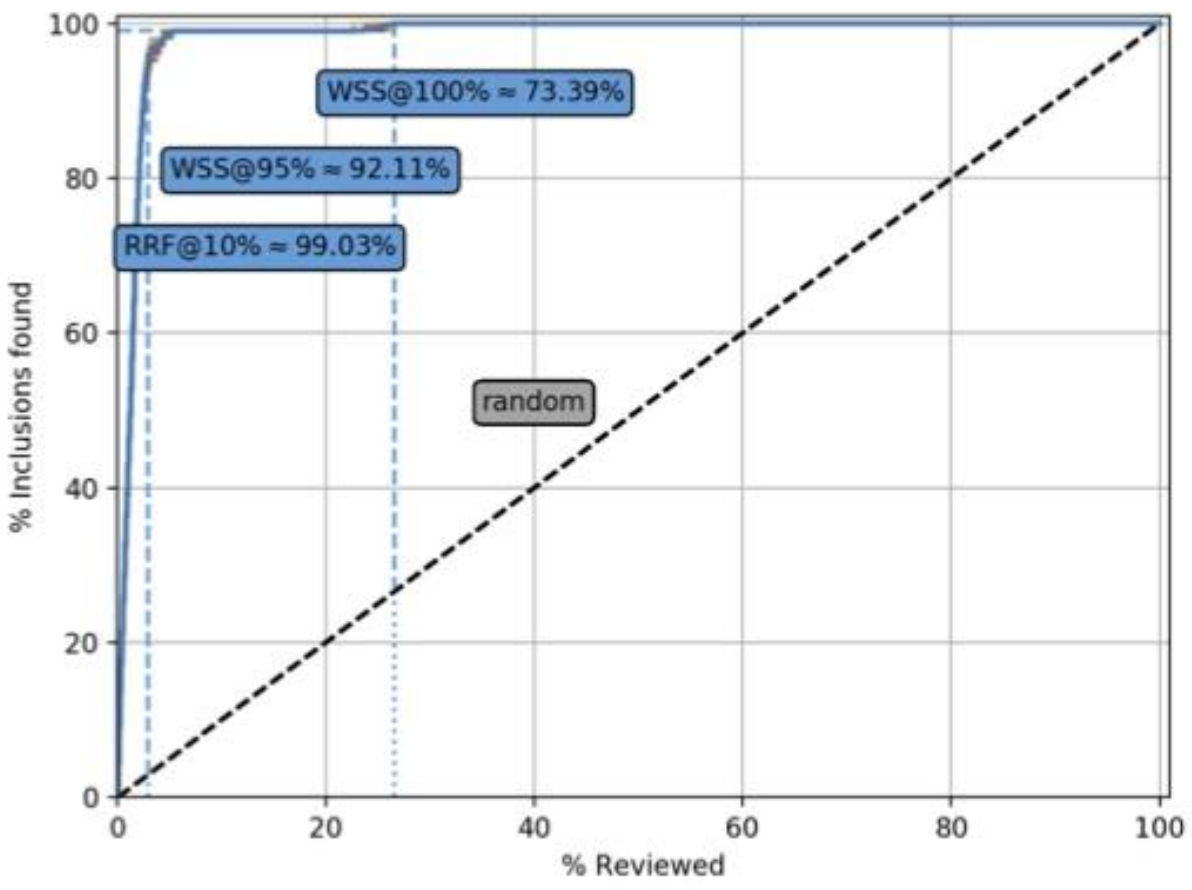

(b)

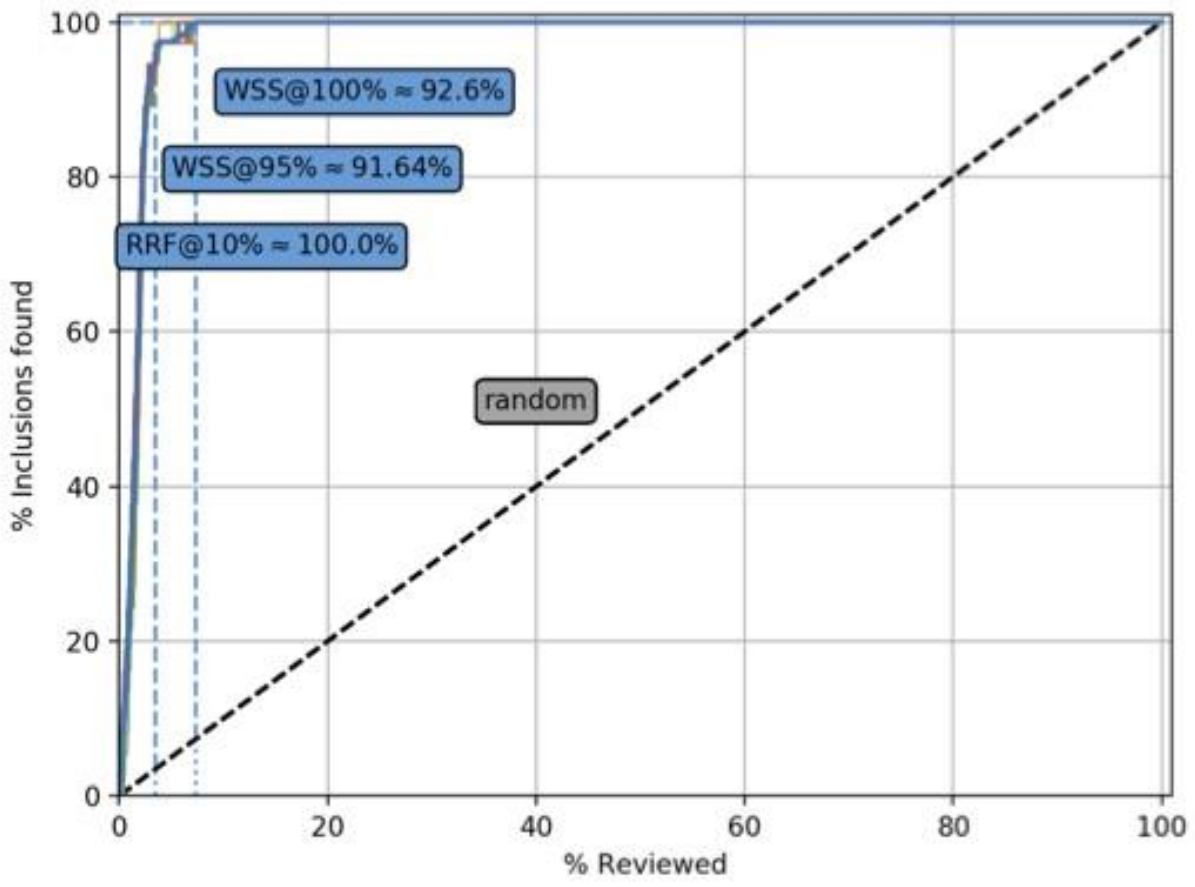

(c)

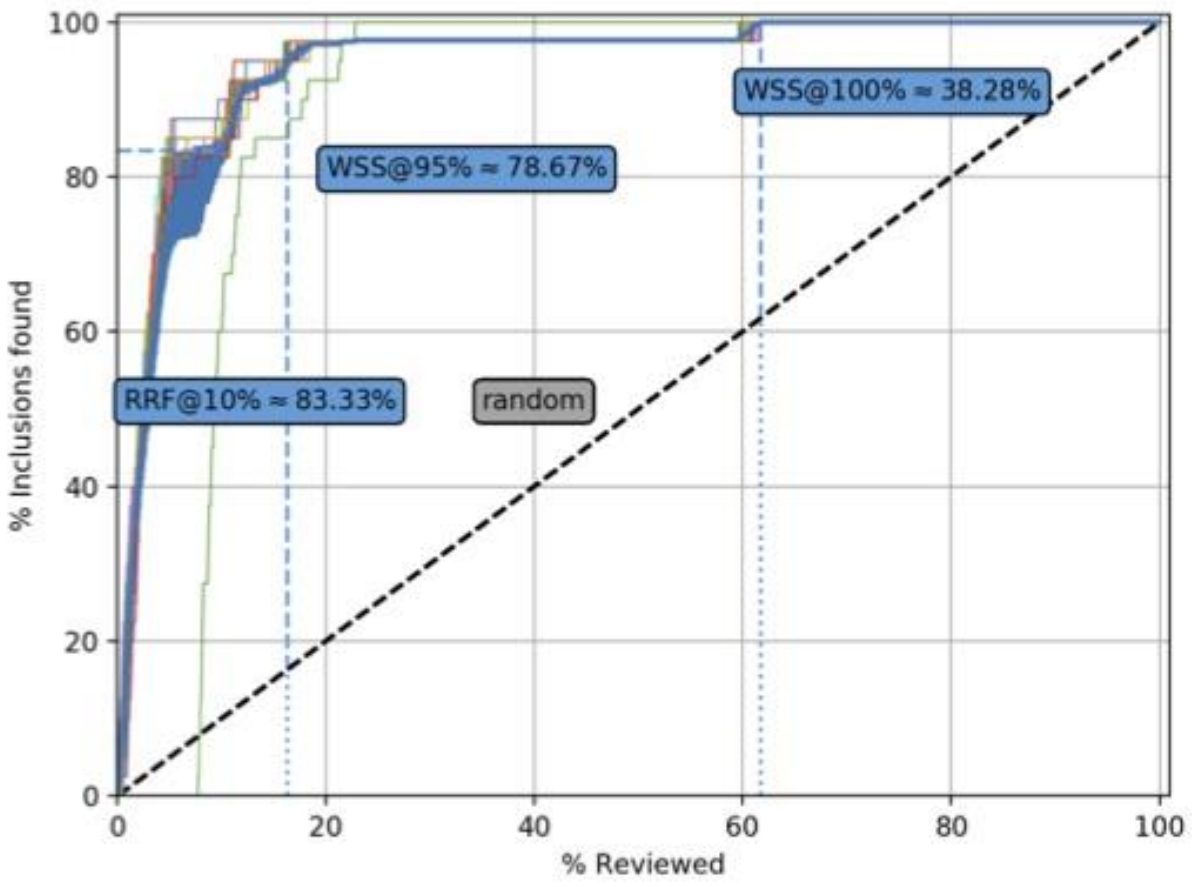

(d)

## References

1. Bornmann, L. & Mutz, R. Growth rates of modern science: A bibliometric analysis based on the number of publications and cited references. *J. Assoc. Inf. Sci. Technol.* **66**, 2215–2222 (2015).
2. Gough, D., Oliver, S. & Thomas, J. *An introduction to systematic reviews*. (Sage, 2017).
3. Cooper, H. *Research Synthesis and Meta-Analysis: A Step-by-Step Approach*. (SAGE Publications, 2015).
4. Liberati, A. *et al.* The PRISMA statement for reporting systematic reviews and meta-analyses of studies that evaluate health care interventions: explanation and elaboration. *J. Clin. Epidemiol.* **62**, e1–e34 (2009).
5. Boaz, A., Ashby, D., Young, K. & others. *Systematic reviews: what have they got to offer evidence based policy and practice?* (ESRC UK Centre for Evidence Based Policy and Practice London, 2002).
6. Oliver, S., Dickson, K. & Bangpan, M. Systematic reviews: making them policy relevant. A briefing for policy makers and systematic reviewers. *EPPI-Cent. Soc. Sci. Res. Unit UCL Inst. Educ. Univ. Coll. Lond. Lond.* (2015).
7. Petticrew, M. Systematic reviews from astronomy to zoology: myths and misconceptions. *Bmj* **322**, 98–101 (2001).
8. Lefebvre, C., Manheimer, E. & Glanville, J. Searching for Studies. in *Cochrane Handbook for Systematic Reviews of Interventions* (eds. Higgins, J. P. & Green, S.) 95–150 (John Wiley & Sons, Ltd, 2008). doi:10.1002/9780470712184.ch6.
9. Sampson, M., Tetzlaff, J. & Urquhart, C. Precision of healthcare systematic review searches in a cross-sectional sample. *Res. Synth. Methods* **2**, 119–125 (2011).

10. Wang, Z., Nayfeh, T., Tetzlaff, J., O'Blenis, P. & Murad, M. H. Error rates of human reviewers during abstract screening in systematic reviews. *PLOS ONE* **15**, e0227742 (2020).
11. Marshall, I. J. & Wallace, B. C. Toward systematic review automation: a practical guide to using machine learning tools in research synthesis. *Syst. Rev.* **8**, 163 (2019).
12. Harrison, H., Griffin, S. J., Kuhn, I. & Usher-Smith, J. A. Software tools to support title and abstract screening for systematic reviews in healthcare: an evaluation. *BMC Med. Res. Methodol.* **20**, 7 (2020).
13. O'Mara-Eves, A., Thomas, J., McNaught, J., Miwa, M. & Ananiadou, S. Using text mining for study identification in systematic reviews: a systematic review of current approaches. *Syst. Rev.* **4**, 5 (2015).
14. Wallace, B. C., Trikalinos, T. A., Lau, J., Brodley, C. & Schmid, C. H. Semi-automated screening of biomedical citations for systematic reviews. *BMC Bioinformatics* **11**, 55 (2010).
15. Cohen, A. M., Hersh, W. R., Peterson, K. & Yen, P.-Y. Reducing Workload in Systematic Review Preparation Using Automated Citation Classification. *J. Am. Med. Inform. Assoc. JAMIA* **13**, 206–219 (2006).
16. Kremer, J., Steenstrup Pedersen, K. & Igel, C. Active learning with support vector machines. *WIREs Data Min. Knowl. Discov.* **4**, 313–326 (2014).
17. Miwa, M., Thomas, J., O'Mara-Eves, A. & Ananiadou, S. Reducing systematic review workload through certainty-based screening. *J. Biomed. Inform.* **51**, 242–253 (2014).
18. Settles, B. *Active Learning Literature Survey*. https://minds.wisconsin.edu/handle/1793/60660 (2009).
19. Holzinger, A. Interactive machine learning for health informatics: when do we need the human-in-the-loop? *Brain Inform.* **3**, 119–131 (2016).
20. Van de Schoot, R. & De Bruin, J. Researcher-in-the-loop for systematic reviewing of text databases. *Zenodo* (2020) doi:10.5281/zenodo.4013207.

21. Kim, D., Seo, D., Cho, S. & Kang, P. Multi-co-training for document classification using various document representations: TF–IDF, LDA, and Doc2Vec. *Inf. Sci.* **477**, 15–29 (2019).
22. Nosek, B. A. *et al.* Promoting an open research culture. *Science* **348**, 1422–1425 (2015).
23. Kilicoglu, H., Demner-Fushman, D., Rindflesch, T. C., Wilczynski, N. L. & Haynes, R. B. Towards Automatic Recognition of Scientifically Rigorous Clinical Research Evidence. *J. Am. Med. Inform. Assoc.* **16**, 25–31 (2009).
24. Gusenbauer, M. & Haddaway, N. R. Which academic search systems are suitable for systematic reviews or meta-analyses? Evaluating retrieval qualities of Google Scholar, PubMed, and 26 other resources. *Res. Synth. Methods* **11**, 181–217 (2020).
25. Borah, R., Brown, A. W., Capers, P. L. & Kaiser, K. A. Analysis of the time and workers needed to conduct systematic reviews of medical interventions using data from the PROSPERO registry. *BMJ Open* **7**, e012545 (2017).
26. de Vries, H., Bekkers, V. & Tummers, L. Innovation in the Public Sector: A Systematic Review and Future Research Agenda. *Public Adm.* **94**, 146–166 (2016).
27. Van de Schoot, R. *et al.* ASReview: Active learning for systematic reviews. *Zenodo* (2020) doi:10.5281/zenodo.3345592.
28. De Bruin, J. *et al. ASReview Software Documentation 0.14*. (Zenodo, 2020). doi:10.5281/zenodo.4287120.
29. ASReview Core Development Team. ASReview PyPI package. https://pypi.org/project/asreview/ (2020).
30. ASReview Core Development Team. Docker container for ASReview. https://hub.docker.com/r/asreview/asreview (2020).
31. Ferdinands, G. *et al.* Active learning for screening prioritization in systematic reviews - A simulation study. *OSF Preprints* (2020) doi:10.31219/osf.io/w6qbg.

32. Fu, J. H. & Lee, S. L. Certainty-Enhanced Active Learning for Improving Imbalanced Data Classification. in *2011 IEEE 11th International Conference on Data Mining Workshops* 405–412 (IEEE, 2011). doi:10/fzbd3k.

33. Le, Q. V. & Mikolov, T. Distributed Representations of Sentences and Documents. *ArXiv14054053 Cs* (2014).

34. Ramos, J. Using TF-IDF to Determine Word Relevance in Document Queries. *Proc. First Instr. Conf. Mach. Learn.* **242**, 133–142 (2003).

35. Pedregosa, F. *et al.* Scikit-learn: Machine Learning in Python. *J. Mach. Learn. Res.* **12**, 2825–2830 (2011).

36. Reimers, N. & Gurevych, I. Sentence-BERT: Sentence Embeddings using Siamese BERT-Networks. *ArXiv190810084 Cs* (2019).

37. Smith, V., Devane, D., Begley, C. M. & Clarke, M. Methodology in conducting a systematic review of systematic reviews of healthcare interventions. *BMC Med. Res. Methodol.* **11**, 15 (2011).

38. Wynants, L. *et al.* Prediction models for diagnosis and prognosis of covid-19: systematic review and critical appraisal. *BMJ* **369**, (2020).

39. Van de Schoot, R. *et al. Extension for COVID-19 related datasets in ASReview*. (Zenodo, 2020). doi:10.5281/zenodo.3891420.

40. Lu Wang, L. *et al.* CORD-19: The Covid-19 Open Research Dataset. *ArXiv ArXiv200410706v2* arXiv:2004.10706v2 (2020).

41. Fraser, N. & Kramer, B. covid19_preprints. (2020) doi:10.6084/m9.figshare.12033672.v18.

42. Ferdinands, G., Schram, R., Van de Schoot, R. & De Bruin, J. Scripts for 'ASReview: Open source software for efficient and transparent active learning for systematic reviews'. *Zenodo* (2020) doi:10.5281/zenodo.4024122.

43. Ferdinands, G., Schram, R., van de Schoot, R. & de Bruin, J. Results for ‘ASReview: Open Source Software for Efficient and Transparent Active Learning for Systematic Reviews’. (2020) doi:10.17605/OSF.IO/2JKD6.
44. Kwok, K. T. T., Nieuwenhuijse, D. F., Phan, M. V. T. & Koopmans, M. P. G. Virus Metagenomics in Farm Animals: A Systematic Review. *Viruses* **12**, 107 (2020).
45. Hall, T., Beecham, S., Bowes, D., Gray, D. & Counsell, S. A Systematic Literature Review on Fault Prediction Performance in Software Engineering. *IEEE Trans. Softw. Eng.* **38**, 1276–1304 (2012).
46. van de Schoot, R., Sijbrandij, M., Winter, S. D., Depaoli, S. & Vermunt, J. K. The GRoLTS-Checklist: Guidelines for reporting on latent trajectory studies. *Struct. Equ. Model. Multidiscip. J.* **24**, 451–467 (2017).
47. van de Schoot, R. *et al.* Bayesian PTSD-Trajectory Analysis with Informed Priors Based on a Systematic Literature Search and Expert Elicitation. *Multivar. Behav. Res.* **53**, 267–291 (2018).
48. Cohen, A. M., Bhupatiraju, R. T. & Hersh, W. R. Feature generation, feature selection, classifiers, and conceptual drift for biomedical document triage. in *TREC* (2004).
49. ASReview Core Development Team. Version 0.4. https://github.com/asreview/asreview/releases/tag/v0.4 (2020).
50. ASReview Core Development Team. Release v0.6. https://github.com/asreview/asreview/releases/tag/v0.6 (2020).
51. ASReview Core Development Team. Release v0.7. https://github.com/asreview/asreview/releases/tag/v0.7 (2020).
52. ASReview Core Development Team. Release v0.9. https://github.com/asreview/asreview/releases/tag/v0.9 (2020).

53. Vasalou, A., Ng, B. D., Wiemer-Hastings, P. & Oshlyansky, L. Human-moderated remote user testing: Protocols and applications. in *8th ERCIM workshop, user interfaces for all, wien, austria* vol. 19 (2004).
54. Joffe, H. Thematic analysis. *Qual. Res. Methods Ment. Health Psychother.* **1**, (2012).
55. QSR International Pty Ltd. *NVivo 12*. (2019).

57. ASReview Core Development Team. Release v0.10. https://github.com/asreview/asreview/releases/tag/v0.10 (2020).
58. ASReview Core Development Team. Release v0.10.1. https://github.com/asreview/asreview/releases/tag/v0.10.1 (2020).
59. ASReview Core Development Team. Release v0.11. https://github.com/asreview/asreview/releases/tag/v0.11 (2020).
60. Marshall, I. J., Kuiper, J. & Wallace, B. C. RobotReviewer: evaluation of a system for automatically assessing bias in clinical trials. *J. Am. Med. Inform. Assoc.* **23**, 193–201 (2016).
61. Nallapati, R., Zhou, B., dos Santos, C., GuÌ‡lçehre, Ç. & Xiang, B. Abstractive Text Summarization using Sequence-to-sequence RNNs and Beyond. in *Proceedings of The 20th SIGNLL Conference on Computational Natural Language Learning* 280–290 (Association for Computational Linguistics, 2016). doi:10/ggvz2s.
62. Xie, Q., Dai, Z., Hovy, E., Luong, M.-T. & Le, Q. V. Unsupervised data augmentation for consistency training. *arXiv preprint arXiv:1904.12848* (2019).
63. Ratner, A. *et al.* Snorkel: rapid training data creation with weak supervision. *VLDB J.* **29**, 709–730 (2020).
64. ASReview Core Development Team. Systematic Review Datasets. https://github.com/asreview/systematic-review-datasets (2020).

65. ASReview Core Development Team. ASReview: Active learning for Systematic Reviews. https://github.com/asreview/asreview (2020).

66. Wallace, B. C., Small, K., Brodley, C. E., Lau, J. & Trikalinos, T. A. Deploying an interactive machine learning system in an evidence-based practice center: abstrackr. in *Proceedings of the 2nd ACM SIGHIT International Health Informatics Symposium* 819–824 (Association for Computing Machinery, 2012). doi:10/fzm9wq.

67. Cheng, S. H. *et al.* Using machine learning to advance synthesis and use of conservation and environmental evidence. *Conserv. Biol.* **32**, 762–764 (2018).

68. Yu, Z., Kraft, N. & Menzies, T. Finding better active learners for faster literature reviews. *Empir. Softw. Eng.* (2018) doi:10/gfr27j.

69. Ouzzani, M., Hammady, H., Fedorowicz, Z. & Elmagarmid, A. Rayyan—a web and mobile app for systematic reviews. *Syst. Rev.* **5**, 210 (2016).

70. Przybyła, P. *et al.* Prioritising references for systematic reviews with RobotAnalyst: A user study. *Res. Synth. Methods* **9**, 470–488 (2018).

## Appendix

### Table A1. Overview of software tools supporting systematic reviews based on Harrison et al. (2020)

| Name | URL | Open Source | Open source URL | Machine Learning | Active Learning |
|---|---|---|---|---|---|
| **Abstrackr**[1] | http://abstrackr.cebm.brown.edu | N | | Y | Y |
| **ASReview**[2] | http://www.asreview.nl/ | Y | https://github.com/asreview/asreview | Y | Y |
| **CADIMA**[3] | https://www.cadima.info/index.php | N | | N | N |
| **Colandr**[4] | https://www.colandrcommunity.com/ | Y | https://github.com/datakind/permanent-colandr-back | Y | Y |
| **Covidence** | https://www.covidence.org/home | N | | N | N |
| **DBPedia**[5] | https://wiki.dbpedia.org/ | N | | Y | Y |
| **DistillerSR** | https://www.evidencepartners.com/ | N | | N | N |
| **EPPI-Reviewer**[6] | https://eppi.ioe.ac.uk/cms/Default.aspx?alias=eppi.ioe.ac.uk/cms/er4 | N | | Y | Y |
| **EROS**[7] | https://www.iecs.org.ar/en/cochrane-centre-at-iecs/eros/ | N | | N | N |
| **FASTREAD**[8,9] | https://github.com/fastread/src | Y | https://github.com/fastread/src | Y | Y |
| **GAPScreener**[10] | https://omictools.com/gapscreener-tool | N | | Y | N |
| **HAWC**[11] | https://hawc.readthedocs.io/en/latest/ | Y | https://github.com/shapiromatron/hawc/tree/master | N | N |
| **JBI-SUMARI**[12] | https://www.jbisumari.org/ | N | | N | N |
| **Lingo 3d**[13] | https://carrotsearch.com/lingo3g/ | N | | Y | N |
| **litstream** | https://www.icf.com/technology/litstream#.Xj0v9_iyPBU.link | N | | N | N |
| **MeSHSIM** | https://github.com/JingZhou2015/MeSHSim | Y | https://github.com/JingZhou2015/MeSHSim | N | N |
| **METAGEAR package for R**[14] | https://cran.r-project.org/web/packages/metagear/index.html | Y | https://github.com/cran/metagear/ | N | N |
| **PARSIFAL** | https://parsif.al/ | Y | https://github.com/vitorfs/parsifal | N | N |
| **PEx**[15] | http://vicg.icmc.usp.br/vicg/tool/1/projection-explorer-pex | N | | N | N |
| **Pimiento**[16] | http://erabaki.ehu.es/jjga/pimiento/ | N | | N | N |
| **Rayyan**[17] | https://rayyan.qcri.org/ | N | | Y | Y |
| **ReLiS**[18] | http://relis.iro.umontreal.ca/auth.html | Y | https://github.com/geodes-sms/relis | N | N |
| **REviewER** | https://sites.google.com/site/eseportal/tools/reviewer | Y | https://github.com/bfsc/reviewer | N | N |
| **ReVis**[19] | ~ | N | | Y | N |
| **RevMan**[20] | https://community.cochrane.org/help/tools-and-software/revman-5 | N | | N | N |
| **revtools**[21] | https://revtools.net/ | Y | https://github.com/mjwestgate/revtools | N | N |
| **RobotAnalyst**[22] | http://nactem.ac.uk/robotanalyst/ | N | | Y | Y |
| **RobotReviewer[23]** | https://www.robotreviewer.net/ | Y | https://github.com/ijmarshall/robotreviewer | Y | N |
| **SESRA** | http://sesra.net/ | N | | N | N |
| **SLR Tool** | http://ta.mdx.ac.uk/slr/about/ | N | | N | N |
| **SLR-Tool[24]** | ~ | N | | Y | N |
| **SLR.qub[25]** | https://github.com/gmergel/SLR.qub | Y | https://github.com/gmergel/SLR.qub | N | N |
| **SLRTOOL[26]** | www.slrtool.org | N | | N | N |
| **SluRp** | https://codefeedback.cs.herts.ac.uk/SLuRp/ | N | | N | N |
| **SRDB.PRO** | https://www.srdb.pro/default | N | | N | N |
| **StArt[27]** | http://lapes.dc.ufscar.br/tools/start_tool | N | | N | N |
| **SWIFT-Active Screener[28]** | https://swift.sciome.com/activescreener | N | | Y | N |
| **SWIFT-Reviewer[29]** | https://www.sciome.com/swift-review/ | N | | Y | N |
| **SyRF** | http://syrf.org.uk/ | N | | N | N |
| **Systematic Review Accelerator[30]** | http://crebp-sra.com/#/ | N | | N | N |

Notes: URL = webpage affiliated with the tool; Open Source = whether source materials of the application are openly available on a collaborative platform such as GitHub; Open Source URL = link to open source code; Machine Learning = whether the tool implements machine learning with the goal of reducing the number of abstracts needed to screen; Active Learning = whether the tool implements machine learning with the goal of reducing the number of abstracts needed to screen; Y=Yes, N=No.

**References Appendix**

1. Gates, A., Johnson, C. & Hartling, L. *Syst. Rev.* **7**, 45 (2018).
2. ASReview Core Development Team. https://doi.org/10.5281/zenodo.3345592 (2019).
3. Kohl, C. *et al. Environ. Evid.* **7**, 8 (2018).
4. Cheng, S. H. *et al. Conserv. Biol.* **32**, 762–764 (2018).
5. Tomassetti, F. C. A. *et al.* in 31–35 (IEE, 2011). doi:10/brzx6d.
6. J. Thomas, J. Brunton & S. Graziosi. (UCL Institute of Education, 2010).
7. Glujovsky, D., Bardach, A., Martí, S. G., Comandé, D. & Ciapponi, A. *Value Health* **14**, A564 (2011).
8. Yu, Z., Kraft, N. & Menzies, T. *Empir. Softw. Eng.* (2018) doi:10/gfr27j.
9. Yu, Z. & Menzies, T. *Expert Syst. Appl.* **120**, 57–71 (2019).
10. Yu, W. *et al. BMC Bioinformatics* **9**, 205 (2008).
11. Shapiro, A., Addington, J., Thacker, S. & Comeaux, J. (Zenodo, 2018). doi:https://doi.org/10.5281/zenodo.1414622.
12. Aromataris, E. & Munn, Z. *Joanna Briggs Inst.* **299**, (2017).
13. Stansfield, C., Thomas, J. & Kavanagh, J. *Res. Synth. Methods* **4**, 230–241 (2013).
14. Lajeunesse, M. J. *Methods Ecol. Evol.* **7**, 323–330 (2016).
15. Joia, P., Coimbra, D., Cuminato, J. A., Paulovich, F. V. & Nonato, L. G. *IEEE Trans. Vis. Comput. Graph.* **17**, 2563–2571 (2011).
16. Garcia Adeva, J. J. & Calvo, R. *IEEE Internet Comput.* **10**, 27–35 (2006).
17. Ouzzani, M., Hammady, H., Fedorowicz, Z. & Elmagarmid, A. *Syst. Rev.* **5**, 210 (2016).
18. Bigendako, B. & Syriani, E. in 552–559 (2020).
19. Felizardo, K. *et al.* in 77–86 (2011). doi:10/fpm52r.
20. (The Cochrane Collaboration, 2014).
21. Westgate, M. J. *Res. Synth. Methods* (2019) doi:10/ggbssh.

22. Przybyła, P. *et al. Res. Synth. Methods* **9**, 470–488 (2018).

23. Marshall, I. J., Kuiper, J. & Wallace, B. C. RobotReviewer: evaluation of a system for automatically assessing bias in clinical trials. J Am Med Inform Assoc 23, 193–201 (2016).

24. Fernández-Sáez, A., Genero, M. & Romero, F. in 157–166 (2010).

25. Mergel, G. D., Silveira, M. S. & da Silva, T. S. in *Proceedings of the 30th annual ACM symposium on applied computing* 1594–1601 (Association for Computing Machinery, 2015). doi:10.1145/2695664.2695902.

26. Barn, B., Raimondi, F., Athiappan, L. & Clark, T. in *Proceedings of the 16th International Conference on Enterprise Information Systems, Volume 2* 440–447 (SCITEPRESS, 2014).

27. Fabbri, S. *et al.* in *Proceedings of the 20th international conference on evaluation and assessment in software engineering* (Association for Computing Machinery, 2016). doi:10/ggf4kg.

28. Miller, K. *et al.* in *Abstracts of the 24th Cochrane Colloquium* (John Wiley & Sons, 2016).

29. Howard, B. E. *et al. Syst. Rev.* **5**, (2016).

30. Cleo, G., Scott, A. M., Islam, F., Julien, B. & Beller, E. *Syst. Rev.* **8**, 145 (2019).